\documentclass[onecolumn,aps,showpacs,superscriptaddress,tightenlines,a4paper]{revtex4}

\usepackage{amssymb}
\usepackage{amsmath}
\usepackage{bbold}
\usepackage{mathrsfs}
\usepackage{paralist}
\usepackage{graphicx}
\usepackage{xcolor}

\begin{document}
%
%
	\title{Ontological aspects of the Casimir Effect}
%
%

	\author{William M. R. Simpson}
	\affiliation{School of Physics and Astronomy, University of St Andrews, North Haugh, St Andrews, KY16 9SS, UK}
	\affiliation{Faculty of Physics, Weizmann Institute of Science, Rehovot, Israel}

%
%
	\begin{abstract}
		The role of the vacuum, in the Casimir Effect, is a matter of some dispute: the Casimir force has been variously described as a phenomenon resulting ``from the alteration, by the boundaries, of the zero-point electromagnetic energy''~\cite{Bordag2001}, or a ``Van der Waals force between the metal plates'' that can be ``computed without reference to zero point energies''~\cite{Jaffe2005}. Neither of these descriptions are grounded in a consistently quantum mechanical treatment of matter interacting with the electromagnetic field. However, the Casimir Effect has been canonically described within the framework of macroscopic quantum electrodynamics \cite{philbin2010}. On this account, the force is seen to arise due to the coupling of fluctuating currents to the zero-point radiation, and it is in this restricted sense that the phenomenon requires the existence of zero-point fields. The conflicting descriptions of the Casimir Effect, on the other hand, appear to arise from inadequate ontologies in which an unwarranted metaphysical priority is assigned either to the matter or the fields. Such ontological errors may have a direct bearing on the problem of the cosmological constant and the correct prediction of the Casimir force in a state of thermal equilibrium.
	\end{abstract}
	%
	%
	\pacs{42.50.-p,42.50.Lc,01.70.+w}
	\maketitle
	\bibliographystyle{unsrt}
%
%

\section{Introduction}

	\par
	The Casimir Effect is an empirically verified quantum mechanical phenomenon involving an attractive force between two parallel uncharged mirrors in vacuum, that exists even at zero temperature~\cite{casimir1948,Lamoreaux1997}. It also attracts the interest of philosophers, as the common explanation for the effect appears to make contact with certain metaphysical categories, such as the ontology of the vacuum~\cite{Saunders2002}. But explanations of the phenomenon are not uniformly consistent among theorists. The Casimir force has been described, on the one hand, as an effect resulting “from the alteration, by the boundaries, of the zero-point electromagnetic energy”~\cite{Bordag2001}. On this account, the force is a property of the vacuum and ``clear evidence for the existence of vacuum fluctuations'' \cite{Carroll2001}. On the other hand, the Casimir Effect has also been described as a ``force [that] originates in the forces between charged particles'' that can be ``computed without reference to zero point energies''. According to this alternative account, ``The Casimir force is simply the (relativistic, retarded) Van der Waals force between the metal plates'' and the phenomenon offers ``no evidence that the zero-point energies are real''~\cite{Jaffe2005}. These descriptions of the Casimir Effect appear to invoke different ontologies in order to account for the phenomenon in question.

Clearly, if the metaphysics of the vacuum is to be informed by the theory of the Casmir Effect, some effort must be made to clarify its requisite ontology. However, popular accounts of the phenomenon involve inadequate ontologies in which an unwarranted metaphysical priority is exchanged between the matter and the fields. Such interpretations are typically grounded in theories which fail to offer a consistently quantum-mechanical description of the interaction of the field with matter. In this author's opinion, the proper locus for interpreting the Casimir Effect is the theory of macroscopic quantum electrodynamics~\footnote{Inevitably, this is an interim position. We know that a better theory will eventually be required because of the deep problems in reconciling quantum field theory and gravity.}, in which the necessary quantization of the electromagnetic field and its coupling to bulk materials receives a canonical and consistently quantum-mechanical treatment. 

%
%
\section{Casimir's formula\label{section-1}}

\subsection{Theoretical context}

In the standard account of the Casimir Effect, the predicted force occurs between a pair of neutral, parallel conducting plates, separated by a distance $d$, in vacuum at zero temperature. The interaction arises due to a disturbance of the vacuum state of the electromagnetic field (in which there are no real photons between the plates)~\cite{casimir1948}. This is a quantum effect, as classical electrodynamics does not predict a force at zero temperature.

The prescribed procedure may be summarised as follows~\cite{Bordag2001}: take the infinite vacuum energy of quantized electromagnetic field, with Dirichlet boundary conditions imposed on the field modes,
\begin{equation}
E =\frac{1}{2} \sum \hbar \omega,
\end{equation}
and subtract from it the infinite vacuum energy in free Minkowski space (or with the boundaries infinitely separated), $E_{\infty}$, having first regularized both quantities $E \rightarrow E(\xi), E_{\infty} \rightarrow E_{\infty}(\xi)$ so that the subtraction procedure is well-defined. Once the difference between the two energies has been computed, the regularization is removed, $\xi \rightarrow 0$, and the result that remains is finite:
\begin{equation}
E_{Casimir} = \text{lim}_{\xi \rightarrow 0} \left[ E(\xi) - E_{\infty}(\xi) \right].
\end{equation}
This is the renormalised Casimir energy, from which we can derive the mechanical force exerted on two parallel plates. For Casimir's case, in which the mirrors are perfectly reflective for all frequencies, we find the pressure force
\begin{equation}
P = - \frac{\hbar c \pi}{240 d^{4}}.
\end{equation}
As an aside, we should observe that nobody follows this recipe exactly for the electromagnetic field, though it has been pedagogically applied to a 1d scalar field~\cite{Bordag2001} where the calculation is somewhat simpler. If we attempt to follow the prescribed procedure precisely, applying a frequency cutoff term $exp(-\xi \omega / c)$ as the regulariser, we discover an additional divergent term that is not removed by subtracting the so-called background energy~\cite{horsleysimpson2013}; it appears to correspond to waves running parallel to the plates. Admittedly, it does not contribute to the force in this case, though it cannot be ignored in other cases~\cite{simpson2013}. Typically it disappears in the course of applying the Euler-McLaurin formula (eg. \cite{leonhardt2010}). Suffice it to say that the simple picture of taking the difference between two energies can be somewhat misleading.

\subsection{Physical interpretation}

Nevertheless, considered on the basis of an energy mode summation, as employed by Casimir (Casimir, 1948), it seems the quantised electromagnetic field in its ground-state, with `external boundary conditions', is sufficient to determine a force -- an almost matter-free prescription for obtaining the phenomenon in which the boundary conditions become simply topological features of the space. Casimir's formula, depending solely upon the the constants $\hbar$ and $c$ and the distance $d$ between the plates, serves to consolidate this impression.

But this interpretation is naive. The vacuum energy, as we have observed, is infinite, and in addition to imposing boundary conditions on the field we must apply some kind of regularisation to tame the mode summation and permit the subtraction of diverging terms. Although the various techniques employed to do this often serve to obscure the fact, it is in the procedure of regularisation that some of the properties of matter (in particular, its dispersive behaviour) are permitted to leak into the calculation, albeit rather crudely~\cite{horsleysimpson2013}. Significantly, it is not possible to extract anything meaningful (or measureable) about the Casimir force until they are permitted to do so. Furthermore, when we relax the highly artificial boundary condition of perfect mirrors, as we must in order to predict the Casimir Effect in real materials, we are forced to sum contributions to the Casimir energy over a dispersive material response across the whole mode spectrum, substantially modifying the predicted force. To do this kind of calculation, however, we must abandon the mode summation and adopt a more sophisticated apparatus, like Lifshitz theory. Casimir's result can still be recovered, but only as a limiting case~\cite{leonhardt2010}.

\section{Lifshitz theory}

\subsection{Theoretical context}

In the context of Lifshitz theory, the Casimir Effect is a result of fluctuating current densities in the two plates~\cite{lifshitz1955,DzyLifPit1961,volume9}. A force arises from the interaction of the currents through the electromagnetic field that they generate in the cavity. The plates are now treated more realistically as dielectric with frequency-dependent permittivities and permeabilities, and this substantially affects both the size (and, in some cases, the nature~\footnote{Lifshitz theory predicts repulsive Casimir forces, under certain circumstances~\cite{capasso2009}.}) of the predicted force.

The formalism is written in terms of the electromagnetic Green function, which describes the field produced by sources of current within the system. A stress tensor is written in terms of this Green function, from which a force can be derived. The stress tensor, however, like the zero-point energy, contains a divergent contribution that must also be regularised~\footnote{Additional divergences in the stress appear in the generalisation to inhomogeneous media (where the optical properties vary continuously along at least one spatial axis). In this case, the regularisation cannot remove the infinities; the spatially dispersive nature of the material must be taken into account~\cite{horsleysimpson2013,simpson2013,leonhardt2011,Simpson2013a}.}. Typically this is achieved through subtracting a stress calculated using an auxiliary Green function associated with an infinite homogeneous medium $\sigma_{0}$ \cite{philbin2011,volume9,philbin2009,leonhardt2010,pitaevskii2011}, and computing the physical stress in the limit of the point of measurement approaching a point source:
\begin{equation}
\sigma_{Casimir} = \text{lim}_{\mathbf{r'} \rightarrow \mathbf{r}} \left[ \sigma(\mathbf{r},\mathbf{r'}) - \sigma_{0}(\mathbf{r},\mathbf{r'}) \right].
\end{equation}
One can then compute a finite stress tensor for the system that depends on the dielectric functions of the material at imaginary frequencies (quantities obtained from the dielectric properties for real frequencies by Hilbert transformation). Only then can the force be derived. Both Casimir's and Lifshitz' regularizations give identical results in the limiting case of a cavity sandwiched between perfectly reflecting mirrors~\cite{leonhardt2010}.

\subsection{Physical interpretation}

An incautious reading of Lifshitz theory might suggest that the role of the vacuum has been successfully banished from the Casimir Effect. Ontologically, the conditions (on first inspection) seem to involve merely the material in the plates and a stochastic source of fluctuations within the material. An electromagnetic field results from the fluctuating currents in the plates, producing a mechanical stress in the same pieces of material that generated it. There is no Hamiltonian in Lifshitz theory and there are no quantized fields. There is therefore no ground-state of the electromagnetic field. To some, this does not even appear to be a quantum-mechanical theory at all~\cite{Knoll2001,Barton2010,philbin2010,philbin2011}.

But the fluctuations in the material that persist even at zero absolute temperature are not a classical phenomenon; they are put in `by hand' using Rytov's correlation function. This can be derived from statistical physics, or from fluctuation-dissipation theorem, and affords the average electromagnetic field that would be present at finite (or zero) temperature~\cite{Rosa2010}.

Lifshitz theory is arguably uncommitted to the particulars of a quantum theory of light in material, however, as opposed to a merely stochastic theory of the phenomena, being based rather on the principles of thermodynamics and statistical physics~\cite{philbin2010,philbin2011}. It embodies at best a minimal treatment of the quantum mechanics that is phenomenologically driven. It is therefore ontologically ambiguous about the role of the vacuum~\footnote{Some ague for the consistency of Lifshitz theory with Casimir's approach. Bordag writes: ``Lifshitz considered the fluctuations in the medium as source. In the modern understanding, these two are equivalent. However, the discussion about two ways continues until present time''~\cite{Bordag2012}. Schwinger, on the other hand, seems to exploit the ambiguity of Lifshitz' theory with his `source theory', replacing the fluctuations of the vacuum with 'source fields' in the plates, with the intention of removing any references to a vacuum state with non-zero physical properties.}.  A clearer interpretation of the underlying Physics cannot be achieved without the vantage point of a more quantum-mechanically consistent position.

\section{Macroscopic QED}

\subsection{Theoretical context}

Macroscopic quantum electrodynamics (macro-QED) offers a canonical quantum-mechanical treatment of the interaction of light with real materials, without the detailed reference to the microscopic material structure that must defeat any complete treatment of such systems, and without sacrificing quantum-mechnical consistency along with more phenomenologically driven approaches~\cite{philbin2010}. It applies with full generality to arbitrary magnetodielectrics, taking full account of the phenomena of dispersion and dissipation.  First, an action is formulated in terms of the dynamical variables $\phi,\mathbf{A}$, the scalar and vector potentials of the fields, and $\mathbf{X}_{\omega},\mathbf{Y}_{\omega}$, a pair of oscillator fields incorporating the dissipative behaviour of the material:
\begin{equation}
S[\phi,\mathbf{A},\mathbf{X}_{\omega},\mathbf{Y}_{\omega}] = S_{em}[\phi, \mathbf{A}] + S_{X}[\mathbf{X}_{\omega}] + S_{Y}[\mathbf{Y}_{\omega}] + S_{int}[\phi,\mathbf{A},\mathbf{X}_{\omega},\mathbf{Y}_{\omega}],
\end{equation}

where $S_{em}$ is the free electromagnetic action, $S_{X}$ and $S_{Y}$ are the actions for the free reservoir oscillators, and $S_{int}$ is the interaction part of the action, coupling the electromagnetic fields to the field reservoirs of the material. Maxwell's equations can be recovered from this action, and canonical quantisation  proceeds straightforwardly. A diagonalised Hamiltonian is achieved,
\begin{equation}
\hat{H} = \frac{1}{2}\sum_{\lambda = e,m}  \int \text{d}^{3}\mathbf{r} \int_{0}^{\infty} \text{d}\omega \, \hbar \omega \, \left( \hat{\textbf{C}}_{\lambda}^{\dagger}(\textbf{r},\omega) \cdot \hat{\textbf{C}}_{\lambda}(\textbf{r},\omega) + \hat{\textbf{C}}_{\lambda}(\textbf{r},\omega) \cdot \hat{\textbf{C}}_{\lambda}^{\dagger}(\textbf{r},\omega) \right),
\end{equation}
where the eigenmodes are bosonic creation and annihilation operators obeying commutation relations
\begin{equation}
\left[ \hat{C}_{\lambda i}(\textbf{r},\omega), \hat{C}_{\lambda ' j}^{\dagger}(\textbf{r} ',\omega ') \right] = \delta_{ij}\delta_{\lambda\lambda '}\delta(\omega-\omega')\,\delta(\textbf{r}-\textbf{r}'),
\quad
\left[ \hat{C}_{\lambda i}(\textbf{r},\omega), \hat{C}_{\lambda ' j}(\textbf{r} ',\omega ') \right] = 0.
\end{equation}
Charge and density operators for the material in the plates, as well as operators for the electromagnetic field, are then expressed in terms of the creation and annihilation operators of the system.
\par
Casimir forces are caused by the stress-energy of the electromagnetic fields in a state of thermal equilibrium. We require the eigenmodes of the system to be in a thermal mixed quantum state. To determine the Casimir force, we consider the ground-state of the system and determine the electromagnetic part of the energy density or stress tensor~\cite{philbin2011} (the complete stress-energy-momentum tensor of the system is obtained from the application of Noether's theorem to the action). The Casimir stress tensor is then the expectation value of the electromagnetic part in thermal equilibrium. At zero-temperature, we recover the zero-point Casimir stress, $\left< \sigma_{ij} \right>$, which has the same form as the more general result for the stress tensor in Lifshitz theory, and from which the Casimir forces in the system can finally be determined, once the stress tensor has been regularised, as before.

\subsection{Physical interpretation}

In macro-QED the materials and the fields are placed on a more equal footing. Physics necessitates the quantization of the fields, and a consistent account of quantised light in material demands a quantum representation of the relevant properties of the material. In the hamiltonian for macro-QED, both aspects of the system are quantised and coupled. The quantum of this system - that is, its irreducible unit of excitation - is a type of polariton.

This in turn means that the matter is coupled to the vacuum state of the fields. Under the condition of thermal equilibrium~\footnote{Lifshitz tensor, commonly used for calculating Casimir forces in realistic systems, is in fact derived under the condition of thermal equilibrim~\cite{lifshitz1955,DzyLifPit1961,pitaevskii2006}.}, it follows that the ground-state of the total system is endowed with physical properties. To see this, consider the following: the fluctuating currents in one plate only interact with the currents in the other if they communicate with them, and they communicate through the electromagnetic field. At zero temperature, there are no photons between the plates, on pain of violating thermodynamic equilibrium; the electromagnetic field is therefore in its ground state. At zero temperature, therefore, the currents in the plates can only communicate through the zero-point radiation.

There is another sense in which macro-QED places the matter and the fields on an equal footing: the  lagrangian formulation that underlies the action, in which the fields, the material and their interaction are posited, is acausal in this respect: we could view the matter as producing the fields, or we could view the fields as inducing the currents in the matter; the actual physics of the phenomenon is indifferent~\footnote{The dynamics are obtained by extremising the action.}.

\section{Three ontologies of the Casimir Effect}

\subsection{Semi-classical ontologies}

Broadly speaking, it is possible to characterise the polarisation of opinion concerning the Casimir Effect into two ontologically distinct positions in which a certain metaphysical priority is exchanged:

\begin{enumerate}[(I).]
  \item a vacuum-field interpretation : there exists an electromagnetic quantum vacuum field, whose properties are modified by material (or topological) constraints, which gives rise to a force.
  \item a Van der Waals material interpretation : there exists a distribution of spontaneously polarised material, whose quantum fluctuations give rise to a force through the retarded electromagnetic field.
\end{enumerate}

The first account touts the existence of a fluctuating electromagnetic quantum field in its ground-state -- the quantum vacuum, a state that is void of particles or quanta, but has the property of an energy available for doing work (for an example of this view, see \cite{Bordag2001}). This property of the field is modified by the imposition of material (or topological) constraints. In the case of the two parallel plates, the energy of the vacuum is reduced (made more negative) by the motion of the plates towards each other. The attractive Casimir force that results is thus a consequence of the zero-point energy of the vacuum; that is, the energy associated with the fluctuations of the vacuum.

On the second account, the Casimir force is essentially reinterpreted as a giant Van der Waals effect (for an example of this view, see \cite{Jaffe2005}). The material in the plates (as opposed to any field between the plates) is subject to quantum fluctuations. These spontaneous disturbances produce field-generating currents within the plates, which interact through the retarded electromagnetic field they have created. These interactions result in a force between the plates.

An essential ontological difference between these two positions is that the first requires a quantum vacuum state with physical properties and the second does not. The Van der Waals interpretation does not assign any physical properties to the vacuum. Importantly, neither of these interpretations is grounded in a consistently quantum-mechanical description of the interaction of the field with matter; the quantum treatment of the problem falls unevenly on one aspect or the other.

\subsection{A dual-aspect ontology}

The position we wish to advance in this paper is distinct:

\begin{enumerate}[(I).]
 \setcounter{enumi}{2}
 \item a dual-aspect-vacuum interpretation :  there exists a vacuum state of the \textit{coupled} system of matter and field, which determines the ground-state properties of the electromagnetic field, giving rise to a force.
\end{enumerate}

On this interpretation, the Casimir force is fundamentally a property of the coupled system of matter and fields, in which the interaction between the plates is mediated by the zero-point fields~\cite{philbin2010,philbin2011}. This interpretation affirms and denies different tenets of interpretations (I) and (II):

\par
First, in common with (I), and in contradiction with (II), there is a vacuum state in (III) which has a physical energy and a role to play in the Casimir Effect. That is, despite the absence of photons between the plates, and thermal vibrations within the plates, the walls of the cavity will still experience an attractive force. Contra (I), but in consort with (II), however, the Casimir Effect does not warrant the assignment of physical meaning to the energy of the vacuum state of the field as a \textit{ding-an-sich}, or totting up its modes in an enormous contribution to the cosmological constant (it is typically cut off at the Planck scale)~\cite{Brand1999,Weinberg1989}. The Casimir Effect offers no justification for quantizing the plane waves of an infinite homogeneous space (which presupposes no coupling to matter) and reifying the `zero-point energy' obtained. There is no force in that case, nor anything for a force to act on. In the quantisation of light coupled to matter, however, the modes of the field are no longer characterised by plane waves. The vacuum state of light is coupled to matter and must be calculated in this context. In other words, it is in this state of interaction that we determine the Casimir energy and measure a Casimir force. Regularisation amounts to drawing a perimeter around this interaction.
\newline
\par
We have described the requisite ontology of the Casimir Effect as `dual-aspect', an appelation that is intended to be sufficiently generous to encompass more detailed accounts of the ontology within the contraints we have discussed. In some sense, the electromagnetic and material aspects of the system are simultaneously present in the vacuum state. However, it is not without the additional structure involved in the details of their interaction that they contribute any actual properties to the system that make contact with observable reality. This irreducible character of the system, in which the `whole' is more than the sum of its `parts', is formally represented in the hamiltonian (via the action) through the addition of an interaction term. In seeking a more detailed account, perhaps we may take our cue from Heisenberg, who saw the wave function as neither the description of any actual state of affairs, nor merely a convenient calculating device, but as referring to a kind of potentiality, and would presumably have described the vacuum similarly~\footnote{Saunders' observation is sapiential. He writes: `on every other of the major schools of thought on the interpretation of quantum mechanics [besides stochastic hidden variable theories]... there is no reason to suppose that the observed properties of the vacuum, when correlations are set up between fields in vacuuo and macroscopic systems, are present in the absence of such correlations'~\cite{Saunders2002}.}. Perhaps for others, the language of emergence may prove the more useful in relating the different aspects of this problem. A more detailed proposal, however, is beyond the scope of this paper.

%
%
	\section{Conclusions}
	\par

In a consistently quantum-mechanical theory of light and matter, the Casimir Effect may be described as a phenomenon arising out of the ground-state properties of a polariton field - a coupled, quantised system of dielectric material and electromagnetic field. In the dual-aspect ground state of the polariton field, the system cannot be separated into material and electromagnetic quanta, since none have been excited. Nevertheless, a Casimir force is predicted between the plates.

In interpreting Casimir's theory, however, the metaphysical emphasis seems to have fallen either on the electromagnetic field or the fluctuating material in the plates. On the first interpretation (I), a universal vacuum field is postulated, in which the ontological role of matter is (perhaps minimally) acknowledged in the form of boundary conditions and the regularisation process. On the second interpretation (II), the bulk material in the plates is prioritised, in which the fluctuating currents generate a field between the cavity that attracts the plates together. The role of the vacuum is void.

However, neither of these interpretations are adequate, and both may lead to false predictions. For a system in thermal equilibrium and at zero temperature, the fields and the currents are in their ground state. Nevertheless, quantum theory predicts a Casimir force. Contra (II), this implies communication through the zero-point fields, and a zero-point energy~\footnote{Lifshitz theory correctly predicts this result, as the stochastic fluctuations in the medium are consistent with the fluctuating field throughout space. However, the Van der Waals interpretation (II) denies that zero-point fields are involved. In thermodynamic equilibrium, at zero temperature, however, it is difficult to see how a zero-force prediction can be avoided, based on this argument, as there are no photons available to communicate between the plates}. In opposition to (I), this does not imply that the energy of the vacuum state in the absence of any coupling is real, nor does it validate the huge contribution to the cosmological constant that this hypostatisation entails.

The Casimir force, ironically, should not be seen through the lens of Casimir's calculation, which is itself without physical application, being at best the limiting case of a more complicated, more abstract, and more consistently quantum-mechanical theory, in which the effects of matter and light are inextricably intertwined.

%
%
	\acknowledgements
	The author wishes to thank T. G. Philbin, S. A. R. Horsley and S. J. Robertson for illuminating discussions, and for critically reading a draft of this paper, as well as SUPA and the Weizmann Institute for their financial support.

\bibliography{refs}

\end{document}